\documentstyle[12pt, manuscript, flushrt, mathptmx]{aastex}

\input epsf
\input psfig.sty

\begin{document}

\sffamily

\title{Chromospheric signatures of small-scale flux emergence as observed with
NST and {\it Hinode} instruments}

\author{
V. B. Yurchyshyn\altaffilmark{1}, P.R. Goode\altaffilmark{1},
V. I. Abramenko\altaffilmark{1},  J. Chae\altaffilmark{1,2},
W. Cao\altaffilmark{1}, A. Andic\altaffilmark{1}, K. Ahn\altaffilmark{1}}

\affil{$^1$ \it Big Bear Solar Observatory, New Jersey Institute of
Technology, Big Bear City, CA 92314, USA}
\affil{$^2$ \it Department of Physics and Astronomy, Seoul National University,
Seoul 151-747, South Korea}

\begin{abstract}

With the ever increasing influx of high resolution images of the solar surface
obtained at a multitude of wavelengths, various processes occurring at
small spatial scales have become a greater focus of our attention. Complex
small-scale magnetic fields have been reported that appear to have enough stored
to heat the chromosphere. While significant progress has been made in
understanding small-scale phenomena, many specifics remain elusive. We present
here a detailed study of a single event of disappearance of a magnetic dipole
and associated chromospheric activity. Based on New Solar Telescope H$\alpha$
data and {\it Hinode} photospheric line-of-sight magnetograms and Ca II H images
we report the following.
1) Our analysis indicates that even very small dipoles (elements separated
by about 0\arcsec.5 or less) may reach the chromosphere and trigger
non-negligible chromospheric activity.
2) Careful consideration of the magnetic environment where the new flux is
deposited may shed light on the details of magnetic flux removal from the solar
surface. We argue that
the apparent collision and disappearance of two opposite polarity elements may
not necessarily indicate their cancellation (i.e., reconnection, emergence
of a ``U'' tube or submergence of $ \Omega $ loops). In our case, the magnetic
dipole disappeared by reconnecting with overlying large-scale inclined plage fields.
3) Bright points seen in off-band H$\alpha$ images are very well-correlated with the
Ca II H bright points, which in turn are co-spatial with G-band bright points. We
further speculate that, in general, H$\alpha$ bright points are expected be co-spatial
with photospheric BPs, however, a direct comparison is needed to refine their
relationship.

\end{abstract}

\section{Introduction}

\citet{fisher1998} showed that X-ray brightness of active regions is related to
the intensity of photospheric magnetic fields. \citet{avi_alex_romano} further
elaborated that more complex and dynamic photospheric fields are usually
associated with the brighter active regions (also see \cite{brooks_2008}).
Although there is no general
agreement as to what mechanisms are responsible for converting magnetic field
energy into heat and transporting it toward the chromosphere and corona, a
functional dependency between the magnetic field and coronal brightness was
implemented in a number of models \citep{lundquist,schrijver2004}.

Only recently, have researchers come to recognize the importance of small-scale
photospheric and chromospheric dynamics for heating and flare activity, e.g.,
\citet{bueno2004}, proposed that the magnetic energy stored in the quiet
photosphere is sufficient to heat the chromosphere. It was also suggested that a
significant part of the heating is generated in the chromosphere in association
with the occurrence of chromospheric jets, which are thought to be a
manifestation of magnetic reconnection or dynamical waves
\citep{shibata2007,root_heating}. Recent {\it Hinode} data show that the
photospheric fields are quite dynamic and complex even on very small scales
\citep{centeno2007,lites2007,orozco}. \cite{brooks_2008} argues that heating
mechanisms of transient events may be substantially  different from that
responsible for bright coronal loops, however they both may be related to highly
dynamic magnetic fields. \citet{ch_parameters} analyzed magnetic
structures in mostly unipolar magnetic settings, such as coronal holes and plage
regions, and concluded that the dynamics of energy release is associated with
small spatial scales, on which magnetic field organization is quite entirely
different. At this moment, there is no generally accepted mechanism to
efficiently convert and transport quiet Sun magnetic energy into the corona. One
of the big problems related to this issue is whether these small-scale fields
(of order of 1\arcsec or smaller) are at all capable of rising high enough to
reach the chromosphere and thus transfer their stored energy. While some
simulations show that emerging $\Omega$-type loops can reach the chromosphere,
other model data indicate that these loops disintegrate at lower heights
\citep{stein2006}. A statistical study of 69 flux emergence events
\citep{martinez2009} concluded that only about 23\% of the emerging dipoles can
reach the chromosphere. Attempts to identify H$\alpha$ arch filament systems or
other signatures associated with the emerging loops have been unsuccessful.

In this case study, we focus on a single event of small-scale magnetic flux
cancellation and associated chromospheric activity. We will show that in mostly
unipolar regions, associated with large-scale fields, even very small loops can
rise high enough to transfer their energy into the chromosphere and corona.

\section{Observations and Data Processing}

New Solar Telescope (NST) in Big Bear Solar Observatory (BBSO) has a 1.7~m
aspheric primary mirror with a 1.6~m clear aperture \citep{goode_nst_2010,
goode_apjl_2010}. NST data used here were obtained at the telescope's Nasmyth
focus with a 0.25\AA~ Zeiss Lyot birefringent H$\alpha$ filter cycling between
H$\alpha$-1.0\AA~ and H$\alpha$+1.0\AA. These off-band blue (red) shifted images
make most apparent upward (downward) directed flows that have velocities up to
50~km/sec (seen as dark features in the respective images). The pixel size of
the images is 0\arcsec.072 and the field of view (FOV) is
67\arcsec$\times$67\arcsec. The original data were speckle reconstructed
employing the KISIP speckle reconstruction code \citep{kisip_code}. Each
reconstructed image was derived from 30 best images selected from a
burst of 100 18~ms exposure images acquired in quick succession.
It is generally recommended to use at least 70 images to successfully perform
speckle reconstruction. However, we found that is cases of moderate seeing image
selection improves contrast of the resulting image. The disadvantage of this approach
is that the high frequency noise may increase, although it is not a concern for this study.
The cadence of the reconstructed images is 8 sec. The resulting speckle
reconstructed images were carefully aligned and destretched.

{\it Hinode} SOT \citep{hinode, tsuneta_sot, suematsu_sot} data utilized here
consist of line-of-sight magnetograms and Ca II H images. The narrowband filter
imager (NFI) operated in a partial resolution mode with a pixel size of 0\arcsec.16, FOV of
225\arcsec$\times$112\arcsec and the magnetic field data were collected on the
Na I 5896\AA~ spectral line with a 3 minute cadence. The data have been
processed with the standard SolarSoft package. The data numbers were then corrected
for polarization sensitivity of the SOT instrumentation as described in
\cite{ichimoto_sot}. The resulting polarization signal was converted
to magnetic flux density adopting weak field approximation as $B\ (Mx\ cm^{-2}) = \alpha * V/I$.
Generally, $ \alpha $ can be determined in several ways, ranging from a simple LTE calculation of
Stokes-V response to the presence of the magnetic field to exploiting an empirical approach and
comparing magnetograms from various instruments. Since there does not exist any
standard calibration routine for \textit{Hinode}/NFI magnetograms, we compared simultaneously
recorded SOT/NFI V/I image and SOT/SP Level2 line of sight magnetic flux and varied parameter
$\alpha$ to produce the most reasonable, in our view, agreement between the two magnetograms.
For this comparison, we used the fraction of the magnetogram's FOV that contained only
plage fields, thus excluding strong sunspot and pore fields. We found that applying $ \alpha$=10000
under-estimates the resulting Na I 5896\AA~ total magnetic flux by about 10
times. When$ \alpha$=23000 the total fluxes calculated from both magnetograms
are equal, which, however, is hardly
acceptable, since we expect the total magnetic field to weaken with height
\citep{abramenko_1992,leka_2003} due to both expansion of the magnetic fields and the the presence
of short magnetic loops which close below the Na I 5896\AA~ line formation level. We therefore,
chose parameter $ \alpha $ based on the following: 1) the chromospheric/photospheric flux ratio
should be about 0.50, and 2) the magnitude of the phosopheric fields should be greater than that of
the chromospheric fields by about 200--500 G \citep{abramenko_1992,leka_2003}.
We thus found that $\alpha $ =16000 best satisfies the above requirements and
this value was accepted as a calibration
parameter. This parameter is two times higher than that used by
\cite{guglielmino}, therefore, we
realize that the produced magnetograms might represent somewhat over-estimated magnitudes of the
magnetic flux, which, however, is  not critical for our study. Variations of $ \alpha $ will
introduce only insignificant changes in the total flux of magnetic elements and the flux change rates.

The Ca II H data came from the {\it Hinode's} broadband filter imager, which operates with a pixel
size of 0\arcsec.11 and a FOV of 190\arcsec$\times$113\arcsec. The NST and {\it Hinode}
Ca II H images were carefully co-aligned by using bright points as a reference to an accuracy of
better than 0\arcsec.25. The correspondence between H$\alpha$ and Ca II H bright points is
discussed in detail in next Section.

Data analyzed in this study were for active region (AR) NOAA 11048 acquired on
2010 February 17. NST observations span from 18:57UT until 20:10UT and overlap
with {\it Hinode's} coverage of this AR. STEREO data revealed that this
slowly evolving active region first appeared on the solar disk at N23W45
on 2010 January 27.

\section{Results}

\begin{figure}
\centerline{\epsfxsize=6.truein  \epsffile{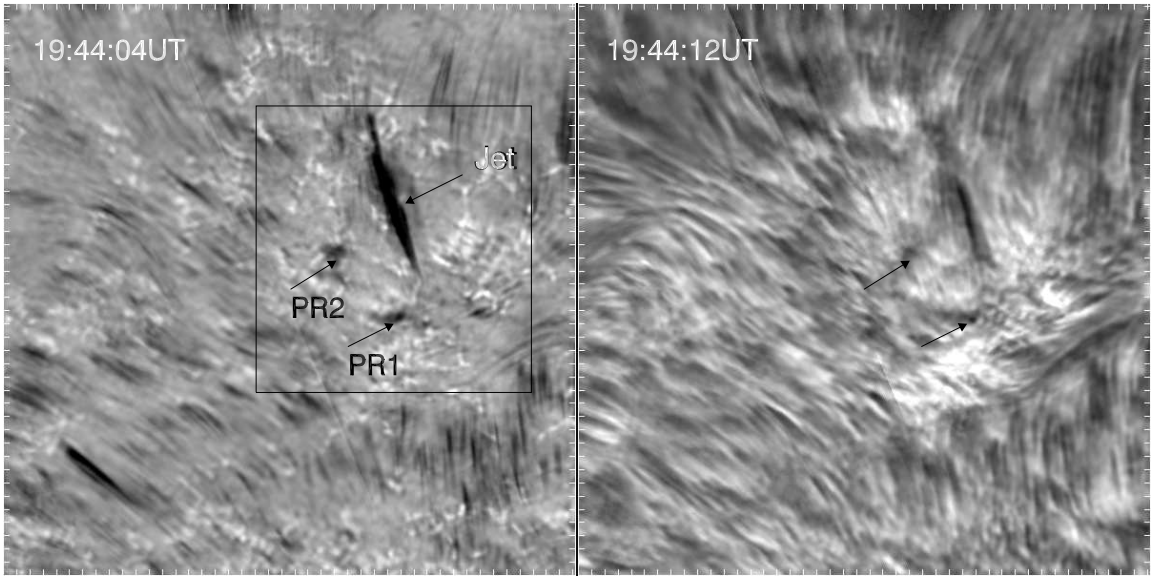}}
\caption{Chromospheric images of AR NOAA 11048 as seen with NST.
Left panel is a 19:44:04UT off-band H$\alpha$-1.0\AA~ image of the trailing part
of the AR, while the right panel shows a nearly simultaneous H$\alpha$ line
center image of the same area. Two arrows indicate two small pores and the box
outlines the region of interest.  The dark jet-like feature seen in the left
panel appears barely discernible in the line center image, which indicates that
nearly all associated plasma move toward the observer with speeds approaching
50km/s.  The tick marks at the edge of the images
indicate 1\arcsec intervals.}
\label{nst}
\end{figure}

Figure \ref{nst} shows NST H$\alpha$-1.0\AA~ (left) and H$\alpha$ line center
(right) images taken at 19:44:04 UT and 19:44:12UT, respectively. The NST FOV
covered only a very small fraction of the trailing part of the AR
dominated by negative polarity fields. The box in Figure \ref{nst} encloses the
area of interest shown in detail in Figure \ref{4panels}. The two upper panels
in Figure \ref{4panels} show a 22\arcsec$\times$22\arcsec portion of the AR
plage, as seen in a H$\alpha$-1.0\AA~ image taken at 19:47:34UT (background).
This image was carefully co-aligned with a 19:47:35UT {\it Hinode}/SOT/FG
magnetogram (contours, left) and a {\it Hinode} Ca II H image acquired at
19:47:32UT (contours, right). Magnetic field contours refer to $\pm$50~G and
$\pm$150~G with red contours representing negative polarity. The lower left
panel shows the same Ca II H image over-plotted with contours of the magnetic
filed. The lower right panel shows H$\alpha$+1.0\AA~ image taken at 19:47:49UT
over-plotted with the 19:47:35UT magnetogram.

The associated chromospheric activity appears to be typical for a plage
region of a quiet AR. There are numerous short lived dark striations associated
with a network area seen against the photospheric background. These striations
seem to be closely associated with the presence of bright points and, according
to \citet{counterparts}, they are rapid blue-shift events thought to be disk
counterparts of type II spicules \citep{bart_two_types}.

The only active dynamic feature seen during the time interval was a dark
recurrent jet-like feature rooted just outside a small pore, PR1. This feature
could be identified at the beginning of the NST observations, however, its
intensity was low. At about 19:30UT, the feature became active and its
visibility in H$\alpha$-1.0\AA~ peaked at about 19:47UT. At this moment, it was
entirely absent in the red-shifted images, while its area and intensity in the
center-line image was much weaker that that in the blue-shifted image.
More precisely, some parts of the feature, mainly the tip of the jet, are not
detectable in the H$\alpha$ center image, while the root of the jet
appears dark at the line center, too). We therefore speculate that the tip
of the jet-like the may be associated with the spectral line shift (i.e.,
mostly plasma up-flows) while its root may display line asymmetry reflecting
a broad spectra of plasma motions. In any case, the fact that the feature is
seen in the far blue wing of the H$\alpha$ spectral line indicates that at
least fraction of the plasma maybe moving at speeds up to 50km/s, which allows
us to identify this feature as a chromospheric jet or H$\alpha$ surge. Numerous
studies of surges \citep[e.g.,][]{chae_surge, yoshimura_surge, achronitis_jet_model}
seem to agree that they are result of magnetic flux cancellation associated with new flux
emergence. Taking into account that the magnetic field, associated with this jet, could be strongly
inclined (as suggested by the presence of dark striations and their well-organized orientations),
we may further speculate that the real velocity of the plasma motion could be even higher.

\subsection{Comparison of Hinode Ca H II and NST H$\alpha$ images}

H$\alpha$-1.0\AA~ images of the lower chromosphere reveal numerous bright points
(BPs) forming long chains and/or small clusters. NST H$\alpha$-1.25\AA~ quiet
Sun data (see Fig. 3 in \citet{goode_apjl_2010}) also showed single BPs located
at vertices of inter-granular lanes. These H$\alpha$ bright points (HBPs) could
be related to the photospheric G-band bright points (GBPs), however, their exact
association is not clear. The present data set does not include G-band images,
however \citet{berger&title2001} compared BPs detected with broadband G-band
(4305\AA) and Ca II K (3933\AA) filters and a narrowband H$\alpha$ filter
(6563\AA). They concluded that large-scale patterns formed by GBPs and magnetic
fields are well-correlated with the GBPs always having underlying magnetic
fields (inverse is not true). Also, co-temporal H$\alpha$-0.7\AA~ images show
increased intensity at the location of the GBPs, however, their correlation
is weaker mainly because H$\alpha$ bright points are masked by numerous dark
absorption features seen in the blue wing of the H$\alpha$ line. \citet{OtsujiPASJ}
reported an association between the G-band and Ca II H bright points (CBPs) with
the CBPs being larger. We, therefore, accept Ca II H bright points as a proxy
for GBPs and address the issue of relationship between the H$\alpha$ bright
points and concentrations of photospheric magnetic fields.

Figure \ref{4panels} (upper right) shows an H$\alpha$-1\AA~ image over-plotted
with contours of Ca II H radiation. Yellow contours indicate areas of enhanced
brightening and dark red contours outline weak Ca II H emission associated with pores
and the jet. The contour levels were set relatively high to avoid crowding the
picture, so some weak CBPs are not plotted. As is evident from these panels,
enhanced brightness in both spectral lines occurs at the locations of strong
magnetic fields  \citep[see also][]{guglielmino}. Also, areas of compact Ca II H
brightenings are co-spatial with enhanced intensity at H$\alpha$ -1.0\AA. Some
HBPs may be at least partially due to enhanced heating caused by the interaction
of magnetic elements in the photosphere. Because Ca II H and G-band BPs are also
well correlated, we may further speculate that in general, H$\alpha$-1.0\AA~ BPs
are expected to be co-spatial with the GBPs, although a detailed analysis would
be beneficial for better understanding of the lower chromosphere of the Sun.

\subsection{Jet Related Dynamics}

\begin{figure}[t]
\centerline{\epsfxsize=5.truein  \epsffile{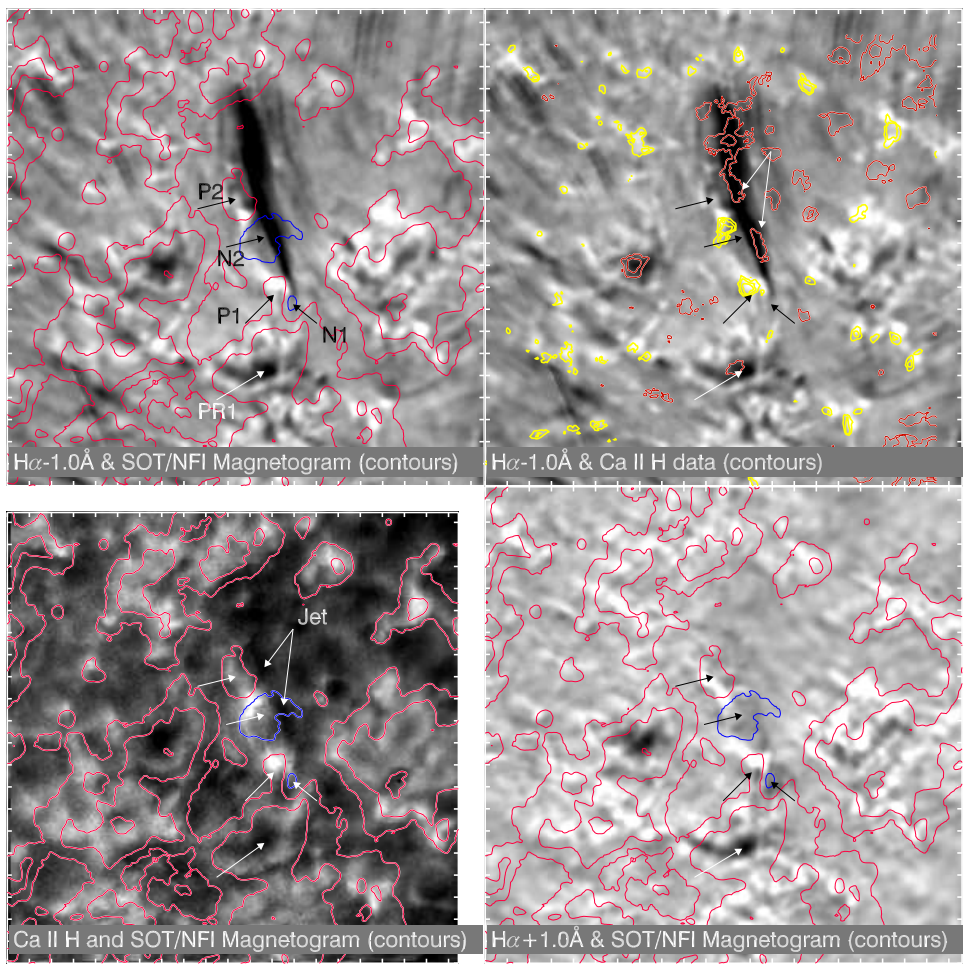}}
\caption{Chromospheric images of AR NOAA 11048 as observed with NST and
{\it Hinode} instruments. The backgrounds in the top panels are the NST
H$\alpha$-1.0\AA~ image (19:47:34UT) of Figure \ref{nst}; in the lower left
panel -- {\it Hinode} Ca II H emission (19:47:32UT) and the lower right panel --
NST H$\alpha$+1.0\AA~ image (19:47:49UT) of Figure \ref{nst}. For comparison,
the upper right image is overplotted with contours of 19:47:32UT {\it Hinode} Ca II H
emission, while the other three images are overplotted with contours of the 19:47:35UT {\it
Hinode}/FG magnetogram. The magnetic field contours are plotted at $\pm$50 and $\pm$150~G levels
with red (blue) contours referring to negative (positive) magnetic polarity. The dark red (yellow)
contours in the upper right panel refer to arbitrary levels chosen to best indicate darkest
(brightest) parts of Ca II H emission without over-crowding the image. The tick marks at the edge
of the images separate 1\arcsec~ intervals. Arrows indicate locations of various features
discussed in the text.}
\label{4panels}
\end{figure}

\begin{figure}
\centerline{\epsfxsize=6.truein  \epsffile{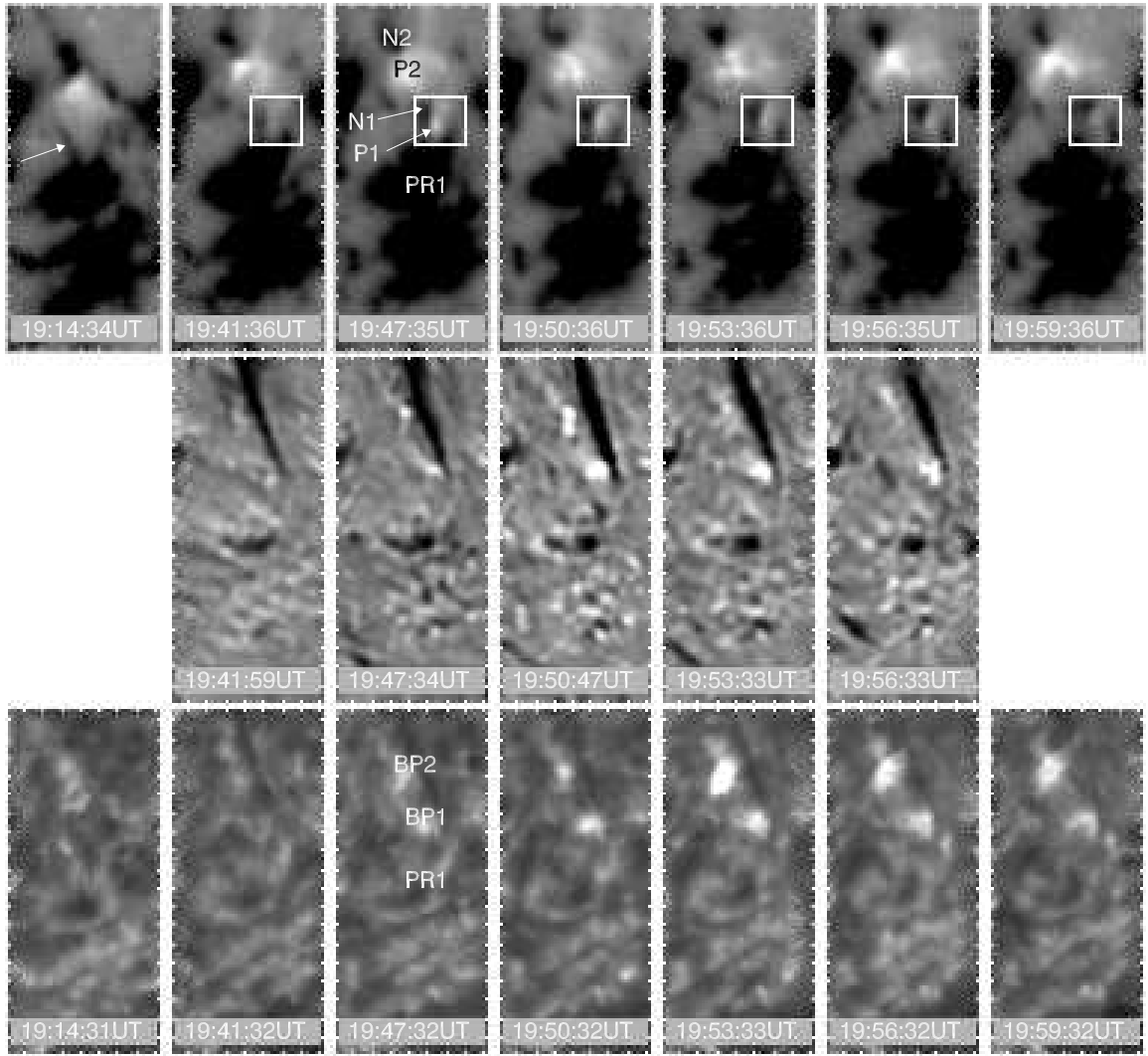}}
\caption{Evolution of the magnetic field surrounding the flux cancellation event
as derived from {\it Hinode}/SOT/FG magnetograms (top), BBSO/NST H$\alpha$-0.75\AA~(middle)
and \textit{Hinode} Ca II H (bottom) images. The data span is nearly one hour. New
mixed polarity flux was constantly appearing in the area between the negative
(dark) and positive (white) polarities. The arrow in the first top panel indicates
the direction of the displacement of negative polarity element, N1. Location of a
small pore is indicated with PR1. P1, N2 and P2 mark locations of magnetic elements
discussed in the text. on The box encloses the area used to measure variations
of magnetic flux. The tick mark separation is 1\arcsec.}
\label{mag}
\end{figure}

As is evident from Figure \ref{4panels}, the H$\alpha$ jet seems to be rooted in
an area dominated by mixed polarity fields, and surrounded by mainly uni-polar,
negative polarity plage-type fields. Figure \ref{mag} shows the evolution of
these mixed polarity fields as seen in a series of {\it Hinode}/SOT/FG
magnetograms spanning nearly one hour. New mixed polarity flux was constantly
appearing in the upper half of the magnetogram. Here we will focus on only one
instance of new flux emergence. Visual inspection of the sequence of the magnetograms
revealed that at about 19:14UT a small, elongated negative polarity magnetic
ridge, N1, began moving toward the center of the image (direction of the
displacement is indicated by the arrow). Estimates yield the average speed of
the ridge to be about 1.5 km s$^{-1}$, which is a typical speed of granular
flows \citep{kubo2010}. As the displacement progressed, a positive polarity
element, P1, began to appear on the right side of the ridge as if the moving
ridge was piling up pre-existing positive fields in front of it. The closest distance
between P1 and N1 centers of gravity measured at 19:50UT was about 0.5~Mm. The
two magnetograms taken before 19:47UT clearly show weak positive fields
``scattered'' in front of the ridge, so that merging (coalescence) of magnetic
fields was indeed possible. Another possibility is that a new magnetic dipole
emerged at this location. Even though there is no solid evidence to support the
merging of a pre-existing flux, this possibility, in general, cannot be excluded
and a non-negligible fraction of newly appearing dipoles may have this character.
Should it be confirmed when it is statistically studied, the merging nature of
new flux appearance may affect our current understanding of flux transport and
visible appearance of through the solar surface.

As it follows from the magnetograms (Figure \ref{mag}), the positive magnetic
flux confined to P1 began to decrease after 19:49UT and it completely
disappeared from the box in Figure \ref{mag} by 20:05UT, while the remnants of
N1 seemed to migrate to the top of the box and eventually left the area. Along
with P1, the positive polarity immediately to the south-west of P1 (lower right
corner of the box) disappeared too. Although P1 and N1 show all the signs of
canceling magnetic features, such as converging motion with subsequent
disappearance, we argue that the line of separation between P1 and N1 may
not be a site of magnetic cancellation, rather, P1 and N1 comprised one magnetic
dipole.

\begin{figure}
\centerline{\epsfxsize=6.truein  \epsffile{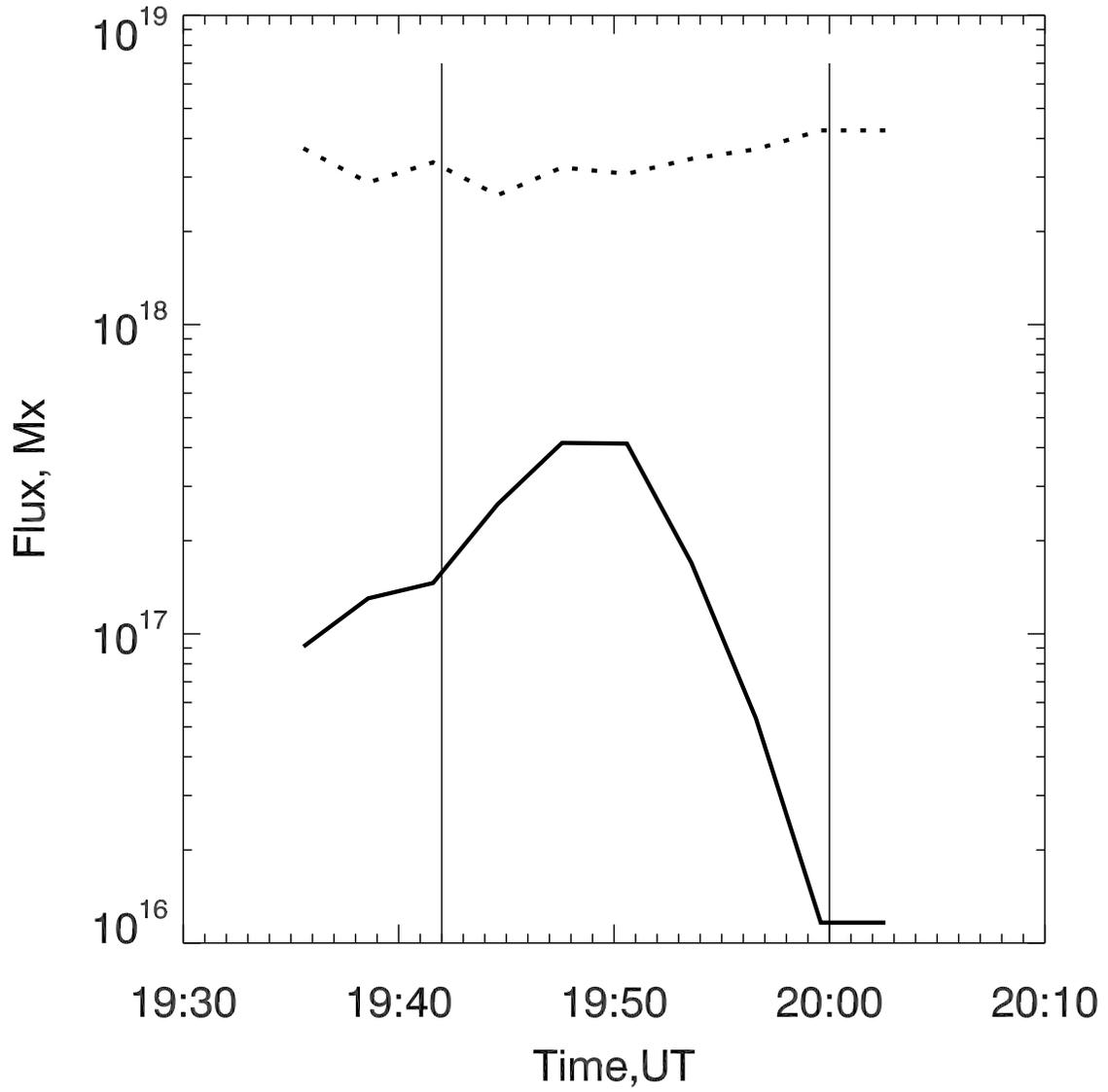}}
\caption{Time profiles of positive (solid) and negative (dashed) magnetic flux
measured inside the box shown in Fig. \ref{mag}. Between 19:50UT and 20:00UT,
the positive magnetic flux decreased by 4.0$\times$10$^{17}$ Mx.}
\label{flux}
\end{figure}

Figure \ref{flux} shows the evolution of the magnetic flux enclosed in the box
in Figure \ref{mag}. Positive polarity flux (solid line) was steadily
increasing and it peaked at about 19:49UT, with the maximum flux density being
about 200~G. The negative polarity flux (dashed) does not show any changes consistent
with positive flux variations, mainly because the corresponding negative
flux changes would be only about 7\% of the pre-existing flux ($6.0\times 10^{18}$ Mx)
and barely discernible in the plot. After 19:49UT, the positive flux began to
decrease and fell by $4.0\times 10^{17}$ Mx over a period of 11 min, which implies that
the magnetic flux changed at a rate of $6.1\times 10^{14}$ Mx s$^{-1}$ (we will call
it the cancellation rate). \citet{chae2002} defined a specific
cancellation rate, which is the cancellation rate divided by the length of the
cancellation interface. If we take the interface to be equal to 0.5~Mm
(full width of P1) the specific cancellation rate is then $1.3\times10^{7}$
Mx$\:$s$^{-1}\:$cm$^{-1}$, which is very similar to the rate of $8\times10^{6}$
Mx$\:$s$^{-1}\:$cm$^{-1}$
reported in \citet{park2009}.

Nearly simultaneously with the enhancement of P1, the dark jet becomes active
and very distinct in blue-shifted H$\alpha$ images (Figure \ref{mag}, middle row), while the co-
temporal Ca II H data (Figure \ref{mag}, bottom row) acquired with a broadband filter showed only
weak signs of the jet. Although not visible in the gray-scale image (Figure \ref{4panels}, lower
left), the Ca II H jet can be distinguished in the contour plot (Figure \ref{4panels}, upper
right), where the blue contours are co-spatial with the H$\alpha$-1.0\AA~ jet. Two bright patches,
(BP1 and BP2) nearly co-spatial with P1 and P2 and seen in off-band H$\alpha$ at 19:47UT, as well
as Ca II H images, similarly revived with enhancement of P1. After 19:50UT they appeared in the Ca
II H images to be connected by bright diffuse structures, which could be interpreted as magnetic
loops connecting magnetic elements P2-N1. Norte that \citet{guglielmino} in Figure 3 show similar
loop like structures connecting a dipole. After 19:50UT the bright patch, BP2, associated with P2
and N2 suddenly brightened and began to rapidly change shape. The shape and brightness evolution
corresponded to that observed during formation and cooling of post-eruptive arcades, although
insufficient spatial resolution precludes us from having solid evidence to support this claim. The
red-shifted H$\alpha$ image over-plotted with magnetic field contours (Figure \ref{4panels}, lower
right) clearly indicates the presence of a well-pronounced dark downflow patch centered at the P1
element. Red-shifted data taken at 20:00 UT also show a dark, high contrast loop connecting the
pore PR1 and P1 fields. These relatively strong downflows could be related either to the plasma
downdrafts along the newly formed loops (similar to downflows in post-eruption arcades) or to the
submergence of these loops.

\begin{figure}
\centerline{\epsfxsize=6.truein  \epsffile{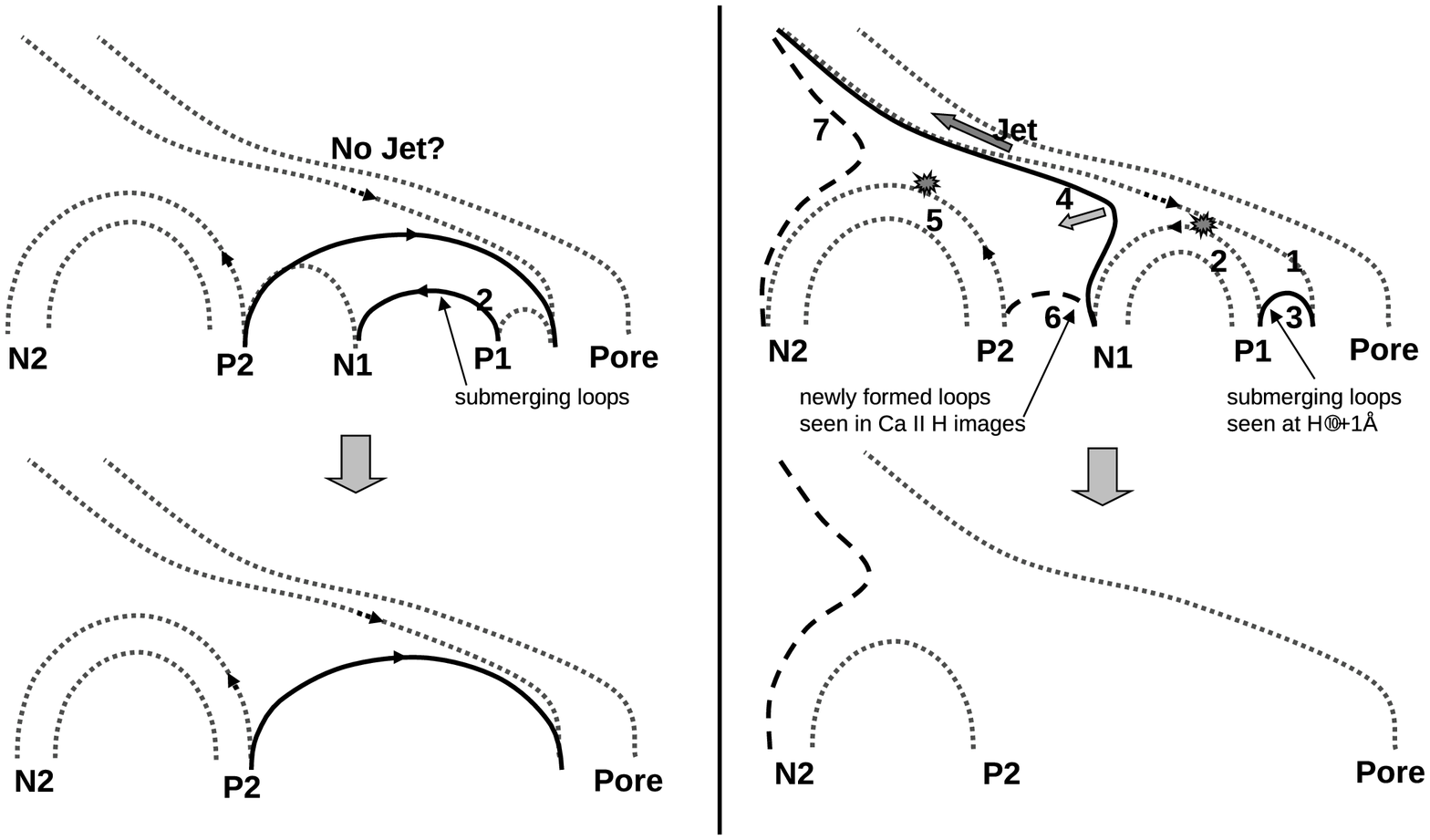}}
\caption{Possible scenarios of flux interaction and reconnection.
In all panels the light gray dotted lines represent magnetic connection before
the cancellation event. Bottom panels show resulting magnetic connections. Left:
P1 and N1 are connected. Lines 3 and 4 represent magnetic connections formed
during first reconnection event between 1 and 2. Lines 6 and 7 are formed due to
the second reconnection event. Right: P1 and N1 belong to different magnetic
systems. Submerging Line 2 is a results of cancellation between P1 and N1.}
\label{scena}
\end{figure}

To interpret the above sequence of events and the dynamics of magnetic
connection during this event, we start with Figure \ref{scena}.
Here we present two possible initial magnetic configurations and the results of
their evolution. First, we assume that magnetic elements P1 and N1 are not
connected and belong to different magnetic systems (Figure \ref{scena}, left).
Please, note that the discussed development is also applicable to the case when
P1 and N1 are footpoints of an emerging ``U''-type loop. The dotted lines
represent magnetic connections before the event. As N1 approaches
P1 they cancel each other via an reconnection event that should occur low in the
chromosphere and photosphere. As a result, two new connections are formed (solid
lines) and P1 and N1 are now connected by a short loop 2, which is expected to
submerge, so that P1 and N1 disappear beneath the solar surface. While this
scenario traditionally treats two approaching magnetic elements as canceling
features, it seems to fail at reproducing the variety of observed features such
as the large-scale jet, the bright Ca II H loops connecting N1 and P2, the
H$\alpha$-1.0\AA~ loop connecting PR1 and P1, activity at N2 and disappearance
of magnetic fields at PR1, P2 and N2.

In the right panels of Figure \ref{scena}, we present an alternative view of
this event. Here we assume that the magnetic elements N1 and P1 comprise one
dipole. Here, too, the dotted lines represent magnetic connections before
the event. As N1 (the negative polarity ridge) moves towards P1 and
enhances it (or as the P1-N1 dipole emerges), magnetic loop 2, connecting them,
expands and eventually reconnects with the opposite directed large-scale fields
1. This interchange reconnection makes line 1 to ``jump'' over the dipole and
a chromospheric jet will be associated with it \citep[e.g.,][]{isobe2008}.
At the same time, element P1 becomes connected to the large area of negative
polarity field associated with pore PR1 (these connecting lines will later be
visible in 20:00UT red-shifted images). As this first reconnection proceeds,
newly created lines 4 will eventually similarly interact with the P2-N2 dipole,
which would lead to another interchange reconnection and formation of a new
connection between P2 and N1 (see diffuse structures connecting P2 and N1 in
the lower left panel of Figure \ref{4panels}). This second reconnection begins
at about 19:50UT, when a chromospheric patch brightens up between P2 and N1.
As a result of this multi-step restructuring, a fraction of large-scale plage
field lines will ``migrate'' away from the pore, while newly created dipoles
P2-N1 and P1-PR1 will fully or partially cancel (submerge).

The latter possibility appears to be more successful in explaining the
observed events. However, in this scenario no cancellation occurs between P1-N1
since it treats the two approaching
magnetic elements as being pre-connected, which is not a commonly accepted
interpretation, although it is quite plausible. This alternative approach may
also explain why instruments fail to detect strong horizontal fields
at many cancellation sites \citep{kubo2010}.

\section{Conclusions and Discussion}

\noindent
1) Our data and analysis show that even very small dipoles (distance between the
elements of order of 0\arcsec.5 or less) may reach the chromosphere and manifest
themselves via non-negligible chromospheric activity.\\
2) Careful consideration of the magnetic environment where the new flux is
deposited may shed light on the details of magnetic flux removal from the solar
surface. We argue that
the apparent collision and disappearance of two opposite polarity elements may
not necessarily indicate their cancellation (i.e., reconnection, emergence
of a ``U'' tube or submergence of $ \Omega $ loops). In our case, the magnetic
dipole disappeared by reconnecting with overlying large-scale inclined plage fields.\\
3) Bright points seen in the off-band H$\alpha$ images are very well-correlated
with the Ca II H bright points, which in turn are co-spatial with G-band
bright points. We further speculate that, in general, H$\alpha$ bright points
expected be co-spatial with photospheric BPs, however, a direct comparison is
needed to refine their relationship.


Small-scale magnetic fields and associated dynamics have recently come to the
center of attention. It was discovered that magnetic flux emerges on very small
scales and it is complex \citep{centeno2007,lites2007,orozco,ch_parameters}.
\citet{bueno2004} and \citet{Ishikawa2008} claim that these small-scale fields
may have enough stored energy to heat the chromosphere. \citet{ch_parameters}
reported that quiet Sun magnetic fields are highly intermittent and complex on
scales below 2~Mm and they may determine the energy release dynamics needed for
chromospheric heating. \citet{shibata2007} observed Ca II H jets in active
regions near the limb and argued that they are evidence of ubiquitous
small-scale reconnection, which may be the source of the chromospheric and
coronal heating. \citet{goode_apjl_2010} recently discovered even tinier
H$\alpha$ jet-like structures in an inter-network field. Thus, the interaction
of the small-scale quiet Sun loops with large-scale fields may be a conduit for
transferring the stored energy up into the chromosphere and corona \citep{martinez2010}.
While some
researchers connect chromospheric jets to reconnection between various magnetic
structures, others argue that a large variety of chromospheric dynamics can be
modeled with a wave-driven reconnection \citep{root_heating}, i.e., small scale
emerging fields may not be essential for chromospheric and coronal energy
balance. Moreover, model calculations also seem to suggest \citep{stein2006}
that small-scale loops cannot reach chromospheric heights, thus deeming this
transport mechanism as not plausible. \citet{martinez2009} reported that about
77\% of loops indeed never rise high enough (or, at least, they were not strong
enough) to show up in chromospheric magnetograms. This case study shows that even
smallest dipoles (0.5~Mm) may interact with the overlying large-scale fields and,
provided that the magnetic environment is favorable (dipole and overlying fields
are oppositely directed), have a significant effect in the chromosphere including
re-structuring of the magnetic field as well as heating and plasma acceleration.

Another issue that our data enables us to addresses is the process of magnetic
flux cancellation. It is accepted now that two magnetic elements of opposite
polarity are canceling if they display converging motions and subsequently
disappear from line-of-sight magnetograms. The current state of this research is
still inconclusive: various studies report that canceling magnetic features may
be associated with both emerging or submerging magnetic fields (see recent paper
by \citet{Iida2010} for more discussion). The direction of the horizontal
magnetic field between two canceling features may help resolve the issue,
however, the 180 degree ambiguity does not allow us to draw any definitive
conclusions. Observed vertical flows at the cancellation site are often used as
an argument to support one view or another, however, these flows may often be
associated with down-drafting plasma along, say, a stationary magnetic loop
rather then with the submergence of an entire loop.

We propose that this problem can benefit from studies of the magnetic
environment where the cancellation occurs. Analysing magnetic connections
before and after the cancellation may lead to more definite conclusions.
We, therefore, argue that although the magnetic dipole in our study displayed
properties of canceling features, it was not a canceling magnetic feature but
rather newly appearing magnetic loops. The flux cancellation occurred as a
result of penetration of the dipole's loops into the chromosphere and their
follow-on interaction with large-scale fields. We also speculate that in this
case, the newly appearing dipole could probably be due enhancement (merging) of an
existing weak flux by converging granular flows, as opposed to an emergence of a
new flux through the solar surface. Similar conclusions come from a statistical
study by \citet{lamb}, who argue that a large fraction of magnetic
flux observed in the quiet Sun could be due to the coalescence of previously
existing week an unresolved flux into concentrations that are large and
strong enough to be detected. This is certainly a possibility that deserves
further exploration.

We thank referee for careful reading of the manuscript and valuable
suggestions and criticism that led to significant improvement of the paper.
Hinode is a Japanese mission developed and launched by
ISAS/JAXA, with NAOJ as domestic partner and NASA and STFC (UK) as international
partners. It is operated by these agencies in co-operation with ESA and  NSC
(Norway). Authors thank BBSO observers and instrument team for their contribution
into this study.
VY work was supported under NASAs GI NNX08AJ20G and LWS TR\&T NNG0-5GN34G
grants. VA acknowledges support from NSF grant ATM-0716512.

\bibliographystyle{/home/arrow/vayur/article/apj}
\bibliography{$HOME/refs,$HOME/my_refs}

\end{document}